\documentclass{aamas2013}

\usepackage{epsfig}
\usepackage{amsmath}
\usepackage{amsfonts}
\usepackage{amssymb}
\usepackage{algorithm}
\usepackage[noend]{algorithmic}
\usepackage{pst-tree}
\usepackage{arydshln}
\usepackage{MnSymbol}
\usepackage{eurosym}
\usepackage{pstricks-add}
\usepackage{multirow}
\usepackage{rotating}

\newtheorem{thm}{Theorem}[section]
\newtheorem{cor}[thm]{Corollary}

\newtheorem{exam}[thm]{Example}

\newtheorem{ques}[thm]{Question}

\begin{document}





\title{On the Verification and Computation of Strong Nash Equilibrium\\ (Extended Version)}

\numberofauthors{3}

\author{
\alignauthor
Nicola Gatti\\
       \affaddr{Politecnico di Milano}\\
       \affaddr{Piazza Leonardo da Vinci 32}\\
       \affaddr{Milano, Italy}\\
       \email{ngatti@elet.polimi.it}
\alignauthor
Marco Rocco\\
       \affaddr{Politecnico di Milano}\\
       \affaddr{Piazza Leonardo da Vinci 32}\\
       \affaddr{Milano, Italy}\\
       \email{mrocco@elet.polimi.it}
\alignauthor
Tuomas Sandholm\\
       \affaddr{Carnegie Mellon}\\
       \affaddr{Computer Science Dep.}\\
       \affaddr{5000 Forbes Avenue}\\
       \affaddr{Pittsburgh, PA 15213, USA}\\
       \email{sandholm@cs.cmu.edu}
}

\maketitle

\begin{abstract}
Computing equilibria of games is a central task in computer science. A large number of results are known for \emph{Nash equilibrium} (NE). However, these can be adopted only when coalitions are not an issue. When instead agents can form coalitions, NE is inadequate and an appropriate solution concept is \emph{strong Nash equilibrium} (SNE). Few computational results are known about SNE. In this paper, we first study the problem of verifying whether a strategy profile is an SNE, showing that the problem is in $\mathcal{P}$.  We then design a spatial branch--and--bound algorithm to find an SNE, and we experimentally evaluate the algorithm.
\end{abstract}









\section{Introduction}
\label{sez:into}

Finding solutions of strategic--form games is a central task in computer science. The most studied solution concept is \emph{Nash equilibrium} (NE)~\cite{shoham-book}. It is appropriate when agents play a game without any form of \emph{a priori} communication, describing a stable state in which no agent can gain more by unilaterally changing  her strategy. Any game has at least a mixed--strategy NE, but searching for it is $\mathcal{PPAD}$--complete~\cite{papastoc} even with two agents~\cite{Chen09:Settling}. It is known that $\mathcal{PPAD} \subseteq \mathcal{NP}$ ($\mathcal{PPAD} \not\subseteq \mathcal{NP}$--complete unless $\mathcal{NP} =\text{co--}\mathcal{NP}$) and it is generally believed that $\mathcal{PPAD} \neq \mathcal{P}$. Thus the worst--case complexity of finding an NE is exponential in the size of the game. Various methods have been adopted to compute NEs, e.g., two--agent games can be solved by linear complementarity mathematical programming (LCP)~\cite{lemke1964}, support enumeration (PNS)~\cite{porter2004}, or mixed--integer linear programming (MILP)~\cite{sandholmgilpinconitzer2005}. 

The NE concept is inadequate when agents can \emph{a priori} communicate, being in a position to form coalitions and deviate multilaterally in a coordinated way. The \emph{strong Nash equilibrium} (SNE) concept strengthens the NE concept by requiring the strategy profile to be resilient also to multilateral deviations, including the grand coalition~\cite{aumann1960}. However, differently from NE, an SNE may not exist. It is known that searching for an SNE is $\mathcal{NP}$--hard, but, except for the case of two--agent symmetric games, it is not known  whether the problem is in $\mathcal{NP}$~\cite{sandholmComplexiy2008}. Furthermore, to the best of our knowledge, there is no algorithm to find an SNE in general games.  The algorithms known in the literature search only for pure--strategy SNEs with specific classes of games, e.g., congestion games~\cite{holzman1997,HayrapetyanSTOC2006,RozenfeldWINE2008}, connection games~\cite{EpsteinACMEC2007}, and maxcut games~\cite{GourvesWINE2009}.

In this paper, we provide a study of verifying and computing an SNE. Our main contributions are as follows.
\begin{itemize}
\item \emph{Verification}. We show that verifying whether a strategy profile in an $n$--agent game is weakly Pareto efficient is in $\mathcal{P}$ (a simple variation of the proof applies to problem of verifying whether a strategy profile in an $n$--agent game is strongly Pareto efficient, showing that also this problem is in $\mathcal{P}$).  We do this by reducing it to the multi--agent minmax problem where the number of actions of the agent whose minmax value is to  be computed is bounded~\cite{peter2008}. Thus, verifying whether a strategy profile is an SNE is in $\mathcal{P}$ and finding an SNE is in $\mathcal{NP}$. The same holds for approximate SNEs.
\item \emph{Computation}. We exploit the verification problem to design an algorithm to search for an SNE. For simplicity, we focus on two--agent games (in principle, our algorithm can be extended to games with more agents) and we design a spatial branch--and--bound algorithm that iterates between the computation of an NE by an oracle and the verification of an SNE. In addition, we show how our algorithm can be improved when the oracle used to find an NE is MIP Nash~\cite{sandholmgilpinconitzer2005} (extending this algorithm to polymatrix games is straightforward).
\item \emph{Experimental evaluation}. 
We show that the ubiquitous benchmark testbed for NE, GAMUT~\cite{gamut}, is not a suitable testbed for SNE because all the instances are easy: if they admit SNEs, then there is always a pure--strategy SNE. Thus, we design a generator whose instances admit only mixed--strategy SNEs. With these instances we experimentally evaluate our algorithms.
\item \emph{Mixed--strategy multilateral deviations}. We show that, differently from what happens with NE, we cannot formulate the conditions of an SNE by a finite set of constraints, one for each pure--strategy uni/multi--lateral deviation.  Rather, mixed--strategy multilateral deviations must be taken into account. This opens the question, left open in this paper, whether or not it is possible to formulate the (necessary and sufficient) conditions of an SNE as a finite set of constraints.
\end{itemize}



\section{Game--theoretic preliminaries}
\label{sez:model}
A strategic--form game is a tuple $(N, A, U)$ where~\cite{shoham-book}:
\begin{itemize}
\item $N=\{1,\dots,n\}$ is the set of agents (we denote by~$i$ a generic agent),
\item $A=\{A_1,\ldots,A_n\}$ is the set of agents' actions and $A_i$ is the set of agent~$i$'s actions (we denote a generic action by $a$,  and by $m_i$ the number of actions in $A_i$),
\item $U=\{U_1\ldots,U_n\}$ is the set of agents' utility arrays where $U_i(a_1,\ldots,a_n)$ is  agent~$i$'s utility when the agents play actions $a_1,\ldots,a_n$.
\end{itemize}
Without loss of generality, $\max_{a_1,\ldots,a_n} \{U_i(a_1,\ldots,a_n)\} = 1$ and $\min_{a_1,\ldots,a_n} \{U_i(a_1,\ldots,a_n)\} = 0$ for every~$i\in N$. We denote by $\mathbf{x}_{i}$ the strategy (vector of probabilities) of agent~$i$ and by $x_{i,a}$ the probability with which agent~$i$ plays action~$a \in A_{i}$. We denote by $\Delta_i$ the space of strategies over $A_i$, i.e., vectors $\mathbf{x}_i$ where the probabilities sum to~1.

The central solution concept in game theory is NE. A strategy profile $\mathbf{x}=(\mathbf{x}_{1},  \ldots, \mathbf{x}_{n})$ is an NE if, for each $i\in N$, $\mathbf{x}_{i}^{T}U_{i}\prod_{j\neq i}\mathbf{x}_{-j} \geq \mathbf{x}_{i}'^{T}U_{i}\prod_{j\neq i}\mathbf{x}_{-j}$ for every $\mathbf{x}_{i}'\in \Delta_i$. Every finite game admits at least an NE in mixed strategies. The problem of finding an NE can be expressed as an NLCP:

\begin{scriptsize}
\begin{align}
\mathbf{x}_i & \geq 0 & \forall i\in N \label{gre_zero}\\
\mathbf{1} v_i - U_{i}\cdot \prod_{j\neq i}\mathbf{x}_{j} & \geq 0 & \forall i\in N \label{exp_utility}\\
\mathbf{x}_i^T \cdot (\mathbf{1}v_i - U_{i}\cdot \prod_{j\neq i}\mathbf{x}_{j}) & = 0 & \forall i\in N \label{compl_conditions}\\
\mathbf{1}^T \cdot \mathbf{x}_i & = 1 & \forall i\in N \label{sum_one}
\end{align}
\end{scriptsize}

\noindent Here $v_{i}$ is the expected utility of agent~$i$.  Constraints~(\ref{gre_zero}) and~(\ref{sum_one}) state that every $\mathbf{x}_i \in \Delta_i$. Constraints~(\ref{exp_utility}) state that no pure strategy of agent~$i$ gives expected utility greater than~$v_i$. Constraints~(\ref{compl_conditions}) state that each agent plays only optimal actions. We denote by $\mathsf{supp}_i$ the support of the strategy of agent~$i$, i.e., the set of actions played with strictly positive probability by~$i$, while $\mathsf{supp}$ denotes the support profile of all the agents. An approximate NE, called $\epsilon$--NE, is a strategy profile~$\mathbf{x}$ in which no agent can improve its utility more than $\varepsilon>0$ by unilaterally deviating.

\begin{figure}[h]
\begin{minipage}{3.5cm}
\vspace{1cm}
\[
\begin{small}
\begin{array}{rr|c|c|}
\multicolumn{2}{c}{}	&	\multicolumn{2}{c}{\textnormal{agent 2}} \\
\multirow{4}{*}{\begin{sideways}agent 1\end{sideways}}	&		&	a_3	&	a_4	\\ \cline{2-4}
	&	a_1	&	3,3 	& 	0,5	\\ \cline{2-4}
	&	a_2	&	5,0	&	1,1	\\ \cline{2-4}
\end{array}
\end{small}
\]
\end{minipage}
\begin{minipage}{5cm}
\begin{pspicture}*(-1,-1)(5.3,5.3)
\scalebox{0.5}{

\psaxes[]{->}(0,0)(0,0)(5.3,5.3)

\psline[linecolor=gray,linewidth=0.25pt]{-}(0,1)(1,1)(1,0)

\psline[linecolor=black,linewidth=2pt]{-}(0,5)(3,3)(5,0)

\psdots[linecolor=black,dotsize=6pt](1,1)

\uput{0}[0](1.2,1.2){NE}

\uput{0}[0](2.1,-0.8){$\mathbb{E}[U_1]$}
\rput[tr]{90}(-0.8,2.9){$\mathbb{E}[U_2]$}
}
\end{pspicture}
\end{minipage}
\caption{Example of game (prisoner's dilemma) without any SNE (left) and Pareto frontier (right).}
\label{exam:noSNE}
\end{figure}
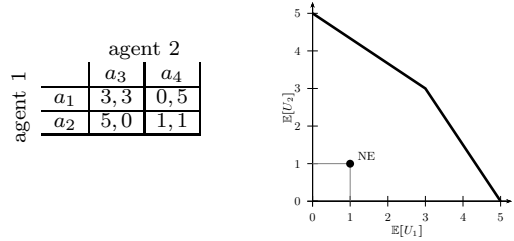

In~\cite{aumann1960}, Aumann introduced the concept of SNE. An SNE strengthens the NE concept requiring the strategy profile to be resilient also to multilateral deviations by any coalition of agents. That is, in an SNE no coalition of agents can deviate in a way that strictly increases the expected utility of each member of the coalition, again keeping the strategies of the agents outside the coalition fixed. An SNE is an NE and it is weakly Pareto efficient for each possible coalition. Differently from NE, SNE is not assured to exist (even in mixed strategies), as shown in the following example.
\begin{exam}
Consider the game  in Fig.~\ref{exam:noSNE}: we report the bimatrix (left) and the Pareto frontier (right). The unique NE is $(a_2,a_4)$, but it is strictly Pareto dominated by $(a_1,a_3)$.
\end{exam}

Finally, an approximate SNE, called $\varepsilon$--SNE, is a strategy profile $\mathbf{x}$ in which no agent can gain more than $\varepsilon>0$ by unilaterally or multilaterally deviating, where the only allowed multilateral deviations are those in which all the members of the coalition strictly improve their utility~\cite{feldman}.

\section{SNE verification}
\label{sec:SNEverification}

Verifying whether a given solution $\overline{\mathbf{x}}$ is an NE is easy for two reasons:
\begin{enumerate}
\item the conditions for a strategy profile to be an NE can be formulated as a set of finite constraints, i.e., (\ref{gre_zero})--(\ref{sum_one});
\item once the strategy variables $\mathbf{x}$ in (\ref{gre_zero})--(\ref{sum_one}) have been assigned values of $\overline{\mathbf{x}}$, the constraints are linear in variables $v_1$ and $v_2$ and then they are easily solvable.
\end{enumerate}
With SNEs it is not clear whether the two above points hold. Although the NE concept requires that a strategy must be the best w.r.t all the mixed strategies, it is sufficient to require that a strategy is the best w.r.t. all the pure strategies. We show that in the case of SNE, this is not sufficient.
For clarity, we focus on two--agent games. In an SNE, in addition to the NE constraints, we need to assure that the agents cannot strictly improve their expected utilities by multilaterally deviating. Limiting the multilateral deviations to pure--strategy deviations is equivalent to requiring that there is no pure outcome that provides both agents with strictly better utilities. We show in the following example that this condition is not sufficient for SNE.

\begin{exam}
Consider the game in Fig.~\ref{exam:SNEnonpure1}. There are three NEs: one pure, $(a_3,a_6)$, and two mixed, $(\frac{1}{2}a_1+\frac{1}{2}a_2,\frac{1}{2}a_4+\frac{1}{2}a_5)$ and $(\frac{1}{7}a_1+\frac{1}{7}a_2+\frac{5}{7}a_3,\frac{1}{7}a_4+\frac{1}{7}a_5+\frac{5}{7}a_6)$. Focus on $(a_3,a_6)$: there is no outcome achievable by pure--strategy multilateral deviations providing both agents a utility strictly greater than $1$. For instance, $(a_1,a_4)$ is better for agent~1 than $(a_3,a_6)$, but it is not for agent~2. With $(a_2,a_4)$ we have the reverse. However, $(a_3,a_6)$ is not weakly Pareto efficient, as shown by the Pareto frontier in the figure. Indeed, $(\frac{1}{2}a_1+\frac{1}{2}a_2,\frac{1}{2}a_4+\frac{1}{2}a_5)$ strictly Pareto dominates $(a_3,a_6)$. Instead, the former being on the Pareto frontier, it is an SNE.
\end{exam}

\begin{figure}[h]
\begin{minipage}{3.5cm}
\vspace{1cm}
\[
\begin{small}
\begin{array}{rr|c|c|c|}
\multicolumn{2}{c}{}	&	\multicolumn{3}{c}{\textnormal{agent 2}} \\
	&		&	a_4	&	a_5	&	a_6	\\ \cline{2-5}
\multirow{3}{*}{\begin{sideways}agent 1\end{sideways}}	&	a_1	&	5,0 	& 	0,5	&	0,0	\\ \cline{2-5}
	&	a_2	&	0,5	&	5,0	&	0,0	\\ \cline{2-5}
	&	a_3	&	0,0	&	0,0	&	1,1	\\ \cline{2-5}
\end{array}
\end{small}
\]
\end{minipage}
\begin{minipage}{5cm}
\begin{pspicture}*(-1,-1)(5.3,5.3)
\scalebox{0.5}{

\psaxes[]{->}(0,0)(0,0)(5.3,5.3)

\psline[linecolor=gray,linewidth=0.25pt]{-}(0,2.5)(2.5,2.5)(2.5,0)

\psline[linecolor=gray,linewidth=0.25pt]{-}(0,1)(1,1)(1,0)

\psline[linecolor=gray,linewidth=0.25pt]{-}(0,0.71)(0.71,0.71)(0.71,0)

\psline[linecolor=black,linewidth=2pt]{-}(0.000000,5.000000)(5.000000,0.000000)

\psdots[linecolor=black,dotsize=6pt](2.5,2.5)(1,1)(0.71,0.71)

\uput{0}[0](2.7,2.7){SNE}

\uput{0}[0](1.2,1.2){NE}

\uput{0}[0](0.125,0.4){NE}

\uput{0}[0](2.1,-0.8){$\mathbb{E}[U_1]$}
\rput[tr]{90}(-0.8,2.9){$\mathbb{E}[U_2]$}
}
\end{pspicture}
\end{minipage}
\caption{Example of two--agent game (left) and its Pareto frontier (right).}
\label{exam:SNEnonpure1}
\end{figure}

We show that formulating the SNE constraints on the basis of only pure--strategy multilateral deviations is not satisfactory even taking into account the sum of the agents' utilities. (The same holds when a generic function of the agents' utilities is employed in place of the sum.)

\begin{exam}
Consider the game in Fig.~\ref{exam:SNEnonpure2}, where $\rho$ is an arbitrarily small positive value. There is a unique NE, $(a_3,a_6)$, and this NE is weakly Pareto efficient being on the Pareto frontier. Notice that the sum of the agents' utilities at $(a_3,a_6)$ is strictly smaller than the sum at $(a_1,a_4)$.
\label{exam:SNEnopure2}
\end{exam}

\begin{figure}[h]
\begin{minipage}{3.5cm}
\vspace{1cm}
\[
\begin{small}
\begin{array}{rr|c|c|c|}
\multicolumn{2}{c}{}	&	\multicolumn{3}{c}{\textnormal{agent 2}} \\
	&		&	a_4	&	a_5	&	a_6	\\ \cline{2-5}
\multirow{3}{*}{\begin{sideways}agent 1\end{sideways}}	&	a_1	&	5,0 	& 	0,0	&	0,\rho	\\ \cline{2-5}
	&	a_2	&	0,0	&	0,5	&	0,\rho	\\ \cline{2-5}
	&	a_3	&	\rho,0	&	\rho,0	&	2,2	\\ \cline{2-5}
\end{array}
\end{small}
\]
\end{minipage}
\begin{minipage}{5cm}
\begin{pspicture}*(-1,-1)(5.3,5.3)
\scalebox{0.5}{

\psaxes[]{->}(0,0)(0,0)(5.3,5.3)
\psline[linecolor=gray,linewidth=0.25pt]{-}(0,2)(2,2)(2,0)

\psline[linecolor=black,linewidth=2pt]{-}(0.000000,5.000000)(0.000500,4.900500)(0.002000,4.802000)(0.004500,4.704500)(0.008000,4.608000)(0.012500,4.512500)(0.018000,4.418000)(0.024500,4.324500)(0.032000,4.232000)(0.040500,4.140500)(0.050000,4.050000)(0.060500,3.960500)(0.072000,3.872000)(0.084500,3.784500)(0.098000,3.698000)(0.112500,3.612500)(0.128000,3.528000)(0.144500,3.444500)(0.162000,3.362000)(0.180500,3.280500)(0.200000,3.200000)(0.220500,3.120500)(0.242000,3.042000)(0.264500,2.964500)(0.288000,2.888000)(0.312500,2.812500)(0.338000,2.738000)(0.364500,2.664500)(0.392000,2.592000)(0.420500,2.520500)(0.450000,2.450000)(0.480500,2.380500)(0.512000,2.312000)(0.544500,2.244500)(0.578000,2.178000)(0.612500,2.112500)(0.648000,2.048000)(0.684500,1.984500)(0.722000,1.922000)(0.760500,1.860500)(0.800000,1.800000)(0.840500,1.740500)(0.882000,1.682000)(0.924500,1.624500)(0.968000,1.568000)(1.012500,1.512500)(1.058000,1.458000)(1.095200,1.433200)(1.125000,1.437500)(1.155200,1.443200)(1.185800,1.450300)(1.216800,1.458800)(1.248200,1.468700)(1.280000,1.480000)(1.312200,1.492700)(1.344800,1.506800)(1.377800,1.522300)(1.411200,1.539200)(1.445000,1.557500)(1.479200,1.577200)(1.513800,1.598300)(1.548800,1.620800)(1.584200,1.644700)(1.620000,1.670000)(1.656200,1.696700)(1.692800,1.724800)(1.729800,1.754300)(1.767200,1.785200)(1.805000,1.817500)(1.843200,1.851200)(1.881800,1.886300)(1.920800,1.922800)(1.960200,1.960700)(2.000000,2.000000)(2.000000,2.000000)(1.960700,1.960200)(1.922800,1.920800)(1.886300,1.881800)(1.851200,1.843200)(1.817500,1.805000)(1.785200,1.767200)(1.754300,1.729800)(1.724800,1.692800)(1.696700,1.656200)(1.670000,1.620000)(1.644700,1.584200)(1.620800,1.548800)(1.598300,1.513800)(1.577200,1.479200)(1.557500,1.445000)(1.539200,1.411200)(1.522300,1.377800)(1.506800,1.344800)(1.492700,1.312200)(1.480000,1.280000)(1.468700,1.248200)(1.458800,1.216800)(1.450300,1.185800)(1.443200,1.155200)(1.437500,1.125000)(1.433200,1.095200)(1.458000,1.058000)(1.512500,1.012500)(1.568000,0.968000)(1.624500,0.924500)(1.682000,0.882000)(1.740500,0.840500)(1.800000,0.800000)(1.860500,0.760500)(1.922000,0.722000)(1.984500,0.684500)(2.048000,0.648000)(2.112500,0.612500)(2.178000,0.578000)(2.244500,0.544500)(2.312000,0.512000)(2.380500,0.480500)(2.450000,0.450000)(2.520500,0.420500)(2.592000,0.392000)(2.664500,0.364500)(2.738000,0.338000)(2.812500,0.312500)(2.888000,0.288000)(2.964500,0.264500)(3.042000,0.242000)(3.120500,0.220500)(3.200000,0.200000)(3.280500,0.180500)(3.362000,0.162000)(3.444500,0.144500)(3.528000,0.128000)(3.612500,0.112500)(3.698000,0.098000)(3.784500,0.084500)(3.872000,0.072000)(3.960500,0.060500)(4.050000,0.050000)(4.140500,0.040500)(4.232000,0.032000)(4.324500,0.024500)(4.418000,0.018000)(4.512500,0.012500)(4.608000,0.008000)(4.704500,0.004500)(4.802000,0.002000)(4.900500,0.000500)(5.000000,0.000000)

\psdots[linecolor=black,dotsize=6pt](2,2)

\uput{0}[0](2.2,2.2){SNE}

\uput{0}[0](2.1,-0.8){$\mathbb{E}[U_1]$}
\rput[tr]{90}(-0.8,2.9){$\mathbb{E}[U_2]$}
}
\end{pspicture}
\end{minipage}
\caption{Example of two--agent game (left) and its Pareto frontier (right).}
\label{exam:SNEnonpure2}
\end{figure}

The above examples show the need for taking into account multilateral deviations in mixed strategies. This could prevent one from formulating the SNE conditions as a finite set of constraints where each constraint is related to a pure--strategy (unilateral/multilateral) deviation, and argues for considering the entire Pareto frontier. Finite constraints expressing the membership of a solution to the Pareto frontier can be derived by using Karush--Kuhn--Tucker conditions~\cite{bertsekas}, but, in the case of non--linear non--convex objective functions as is the case with SNE, the set of constraints is only necessary and not sufficient (and therefore we have no guarantee that a strategy is on the Pareto frontier). Nevertheless, although it is not clear whether one can formulate the SNE problem with a set of finite constraints as in the case of NE, we can show that the SNE verification problem is tractable.

We first study the problem of verifying whether a given solution is weakly Pareto efficient.
\begin{thm}
Given an $n$--player game with rational payoffs and a strategy profile $\mathbf{x} = (\mathbf{x}_1, \ldots, \mathbf{x}_n)$ with rational probabilities, the problem of verifying whether $\mathbf{x}$ is weakly Pareto efficient is in~$\mathcal{P}$ when $n$ is constant.
\label{thm:efficiency}
\end{thm}
\emph{Proof}.
Recall that $\mathbf{x}$ is weakly Pareto efficient if

\begin{scriptsize}
\begin{multline}
\not \exists \mathbf{x}'=(\mathbf{x}'_1, \ldots, \mathbf{x}'_n), \mathbf{x}'_i\in \Delta_i,\not \exists \phi>0: \\
\forall i\in N, \mathbf{x}_i' \cdot U_i \cdot \prod_{j\neq i} \mathbf{x}_j' - \mathbf{x}_i \cdot U_i \cdot \prod_{j\neq i} \mathbf{x}_j \geq \phi \label{def:weakPareto}
\end{multline}
\end{scriptsize}

\noindent The structure of the proof of theorem is the following:
\begin{itemize}
\item (\emph{Step 1}) we show that the problem of verifying whether $\mathbf{x}$ is weakly Pareto efficient is equivalent to the problem of verifying whether the minmax value of a fictitious agent vs. the set $N$ of agents is non--negative;
\item (\emph{Step 2}) we show that the problem of verifying whether the minmax value of a fictitious agent vs. the set $N$ of agents is non--negative is in $\mathcal{P}$.
\end{itemize}
Next we discuss these two steps in detail.

\emph{Step 1}. Focus on $\mathbf{x}_i \cdot U_i \cdot \prod_{j\neq i} \mathbf{x}_j$: it is a rational value expressing the expected utility of agent~$i$ given strategy profile~$\mathbf{x}$. Call $\tilde{U}_i = M_1 \cdot (\mathbf{x}_i \cdot  U_i \cdot  \prod_{j\neq i} \mathbf{x}_j)- U_i$, where $M_1$ is a multidimensional array of ones. We can show that

\begin{scriptsize}
\begin{align*}
\mathbf{x}_i' \cdot \tilde{U}_i \cdot \prod_{j\neq i} \mathbf{x}_j' & = \mathbf{x}_i' \cdot M_1 \cdot (\mathbf{x}_i \cdot U_i \cdot \prod_{j\neq i} \mathbf{x}_j) \cdot \prod_{j\neq i} \mathbf{x}_j' - \mathbf{x}_i' \cdot U_i \cdot \prod_{j\neq i} \mathbf{x}_j'\\
& = (\mathbf{x}_i \cdot U_i \cdot \prod_{j\neq i} \mathbf{x}_j) \cdot \mathbf{x}_i' \cdot M_1 \cdot  \prod_{j\neq i} \mathbf{x}_j' - \mathbf{x}_i' \cdot U_i \cdot \prod_{j\neq i} \mathbf{x}_j'\\
& = \mathbf{x}_i \cdot U_i \cdot \prod_{j\neq i} \mathbf{x}_j - \mathbf{x}_i' \cdot U_i \cdot \prod_{j\neq i} \mathbf{x}_j'
\end{align*}
\end{scriptsize}

\noindent given that $\mathbf{x}_i' \cdot M_1 \cdot  \prod_{j\neq i} \mathbf{x}_j' =1$ because it is a (non--convex) combination of ones. Thus, we can write the condition  $\mathbf{x}_i' \cdot U_i \cdot \prod_{j\neq i} \mathbf{x}_j' - \mathbf{x}_i \cdot U_i \cdot \prod_{j\neq i} \mathbf{x}_j \geq \phi$ in~(\ref{def:weakPareto})  as $\mathbf{x}_i' \cdot \tilde{U}_i \cdot \prod_{j\neq i} \mathbf{x}_j'  \leq -\phi$.

Now, we consider the following optimization problem

\begin{scriptsize}
\begin{align}
\min_{\gamma, \mathbf{x}_i'~i\in N}\qquad	& \gamma		\label{eq:minmaxN1} \\
						& \gamma						&& \geq \mathbf{x}_i' \cdot \tilde{U}_i \cdot \prod_{j\neq i} \mathbf{x}_j'	& \forall i\in N	\label{eq:minmaxN2} \\
						& \mathbf{1}^T\cdot \mathbf{x}'_i	&& = 1		& \forall i\in N							 \label{eq:minmaxN3} \\
						& \mathbf{x}'_i					&& \geq \mathbf{0} & \forall i\in N						 \label{eq:minmaxN4}
\end{align}
\end{scriptsize}

\noindent This minimizes the value of~$\gamma$ subject to the constraint that $\gamma$ cannot be smaller than $\mathbf{x}_i' \cdot \tilde{U}_i \cdot \prod_{j\neq i} \mathbf{x}_j'$ for every~$i \in N$. If the optimal value~$\gamma^*$ is strictly smaller than zero, then $\mathbf{x}_i' \cdot \tilde{U}_i \cdot \prod_{j\neq i} \mathbf{x}_j'<0$ for every~$i\in N$ and therefore there exists $\phi> 0$ such that $\mathbf{x}_i' \cdot \tilde{U}_i \cdot \prod_{j\neq i} \mathbf{x}_j'  \leq -\phi$. That is, if $\gamma^*<0$, then $\mathbf{x}$ is strongly Pareto dominated. Instead, if~$\gamma^*=0$ (it can be observed that $\gamma^*$ is never strictly positive), we cannot find such a $\phi$ and therefore $\mathbf{x}$ is weakly Pareto efficient.

Problem~(\ref{eq:minmaxN1})--(\ref{eq:minmaxN4}) can be interpreted as the minmax problem in which all the agents $i\in N$ minimize (without correlation) the maximum value of a fictitious agent $i_f$ (the $n+1$--th): the fictitious agent $i_f$ has~$n$ actions and each action~$i$ provides an expected utility of $\mathbf{x}_i' \cdot \tilde{U}_i \cdot \prod_{j\neq i} \mathbf{x}_j'$.

\emph{Step 2}. Here we consider the problem of verifying whether the minmax value of an agent with $n$ available actions against $n$ opponents, each agent~$i$ with $m_i=m$ actions, is (strictly) smaller than a value (in our case zero) when $n$ is given. In~\cite{peter2008}, the authors show that such a problem can be solved in time $L^{O(1)} \cdot k_1^{O(k_1\cdot k_2 )} \cdot k_3^{k_1 \cdot k_2}$ where:
\begin{itemize}
\item $L$ is the bit complexity of the game;
\item $k_1$ is the total number of agents (in our case, $k_1 = n+1$);
\item $k_2$ is the number of actions available to the agent whose minmax value is to be computed (in our case, $k_2=n$);
\item $k_3$ is the number of actions available to the other agents (in our case, $k_3=m$).
\end{itemize}
As a result, we have $L^{O(1)} \cdot (n+1)^{O(n^2)} \cdot m^{n^2+n}$. Given that in our problem $k_1=k_2=n$ and $n$ is fixed, the computational complexity is polynomial in the size of the game. To complete the proof of the theorem, we need to show even when each agent~$i\in N$ has a potentially different number of actions $m_i$, the computational complexity stays polynomial. Let $\overline{m}=\max_{i\in N}\{m_i\}$ be the maximum number of actions a (non--fictitious) agent can have, and let $\overline{i}=\arg\max_{i\in N}\{m_i\}$. For each agent~$i\in N, i\neq \overline{i}$, we can add $\overline{m}-m_i$ fictitious actions and we can associate each outcomes induced by fictitious actions with $1$ for the fictitious agent (recall from Section~\ref{sez:model} that the maximum utility of each agent is 1). In this way, each agent~$i\in N$ has the same number $\overline{m}$ of actions and the extra actions do not affect the computation of the minmax value, agents never playing the extra ones.
\hfill$\Box$

The algorithm for verifying whether strategy profile $\mathbf{x}$ is weakly Pareto efficient can be obtained by adapting the algorithm presented in~\cite{peter2008} to our problem. It is reported in Algorithm~\ref{alg:Pareto} where

\begin{scriptsize}
\begin{multline*}
\hspace{-0.3cm}\psi(N,\mathbf{x},\mathsf{supp})=\\
\hspace{0.1cm}\left\{
\begin{array}{lr}
\bigwedge_{i\in N} \left(\sum_{a_1\in \mathsf{supp}_1}\ldots\sum_{a_n\in \mathsf{supp}_n} (\tilde{U}_i(a_1,\ldots,a_n)\cdot \prod_{k\in N}x'_{k,a_{k}}<0)\right) &\hspace{-0.2cm}\wedge\\
\bigwedge_{i\in N} \left(\sum_{a_i\in \mathsf{supp}_i}x'_{i,a_i}=1\right) &\hspace{-0.2cm}\wedge\\
\bigwedge_{i\in N, a_i\in \mathsf{supp}_i} \left(x'_{i,a_i}\geq 0\right)
\end{array}\right.
\end{multline*}
\end{scriptsize}

\noindent The algorithm enumerates (by means of $\mathsf{enumerate}$) all the possible joint supports of size $n$, because, as shown in~\cite{peter2008}, there always exists a minmax strategy for the problem~(\ref{eq:minmaxN1})--(\ref{eq:minmaxN4}) in which all the agents $i\in N$ randomize over at most a number of actions equal to the number of actions of the agent whose minmax value is to be computed  (in our case, $n$). Thus we need to enumerate $\prod_{i\in N}\binom{m_i}{n}$ different joint supports of size $n$. For each joint support, we need to evaluate formula $\psi$ and we can do that by using the algorithm presented in~\cite{basu}. If $\psi$ is true, then there is a strategy profile $\mathbf{x}'$ that Pareto dominates the input $\mathbf{x}$.

\begin{algorithm}
\begin{algorithmic}[1]
\STATE $E\longleftarrow \mathsf{enumerate}(\mathsf{supp}_1,\ldots,\mathsf{supp}_n:\forall i,|\mathsf{supp}_i|=n)$
\FORALL{elements of $E$}
        	\IF {$\exists (\mathbf{x}'_1,\ldots, \mathbf{x}'_n)\in \mathbb{R}^{n^2}:[\psi(N,\mathbf{x},\mathsf{supp})]$}
		\RETURN $(\mathsf{false},\mathbf{x}')$
	\ENDIF
\ENDFOR
\RETURN $(\mathsf{true},\emptyset)$
\end{algorithmic}
\caption{$\mathsf{verifyPareto}(N,\{U_i\}_{i\in N},\mathbf{x})$}
\label{alg:Pareto}
\end{algorithm}

We leverage Theorem~\ref{thm:efficiency} to state the following.

\begin{thm}
Given an $n$--player game with rational payoffs and a strategy profile $\mathbf{x} = (\mathbf{x}_1, \ldots, \mathbf{x}_n)$ with rational probabilities, verifying whether $\mathbf{x}$ is an SNE is in~$\mathcal{P}$ when $n$ is constant.
\label{thm:verificationSNE}
\end{thm}
\emph{Proof}.
Strategy profile $\mathbf{x}$ is an SNE if:
\begin{itemize}
\item (\emph{Condition 1}) it is an NE,
\item (\emph{Condition 2}) for every $N'\subseteq N$, $(\mathbf{x}_{i}:i\in N')$ is weakly Pareto efficient once the strategies $\mathbf{x}_j$ of all the other agents~$j\in N\setminus N'$ are fixed.
\end{itemize}
Checking Condition~1 is easy; it is checking whether or not a polynomial number of constraints is satisfied. Checking Condition~2 requires one to verify whether  $O(2^n)$ strategy profiles are weakly Pareto efficient where $n$ is given. Checking weak Pareto efficiency is easy as Theorem~\ref{thm:efficiency} shows and the number of verification problems is a constant in the size of the problem, $n$ being fixed. 
\hfill$\Box$

The algorithm to verify whether $(\mathbf{x}_1,\ldots,\mathbf{x}_n)$ is an SNE is reported in Algorithm~\ref{alg:SNEverification}. It enumerates all the possible coalitions of agents (except the empty coalition), and for each coalition it verifies whether the coalition strategy is weakly Pareto efficient. If it is not, the algorithm returns the coalition $N'$ for which a multilateral deviation is possible and a strongly Pareto dominant coalition strategy $\mathbf{x}'$ (notice that $\mathbf{x}'$ is not assured to be weakly Pareto efficient).

\begin{algorithm}
\begin{algorithmic}[1]
\STATE $E\longleftarrow \mathsf{enumerate}(N'\subseteq N: N\neq \emptyset)$
\FORALL{element of $E$}
	\STATE $\mathbf{x}''=(\mathbf{x}_i)_{i \in N'}$
	\STATE $(\vartheta,\mathbf{x}')\longleftarrow \mathsf{verifyPareto}(N',\{ U_i\prod_{j\not \in N'}\mathbf{x}_j\}_{i \in N'},\mathbf{x}'')$
        	\IF {$\neg \vartheta$}
		\RETURN $(\mathsf{false},N',\mathbf{x}')$
	\ENDIF
\ENDFOR
\RETURN $(\mathsf{true},\emptyset,\emptyset)$
\end{algorithmic}
\caption{$\mathsf{verifySNE}(\mathbf{x})$}
\label{alg:SNEverification}
\end{algorithm}

We can show that the previous theorem holds even when we focus on strong Pareto efficiency. More precisely, we can state what follows.
\begin{thm}
Given an $n$--player game with rational payoffs and a strategy profile $\mathbf{x} = (\mathbf{x}_1, \ldots, \mathbf{x}_n)$ with rational probabilities, the problem of verifying whether $\mathbf{x}$ is strongly Pareto efficient is in~$\mathcal{P}$ when $n$ is constant.
\label{thm:efficiencystrong}
\end{thm}
\emph{Proof}. The proof is a variation of the proof of Theorem~\ref{thm:efficiency}.
Recall that $\mathbf{x}$ is strongly Pareto efficient if

\begin{scriptsize}
\begin{multline}
\not \exists \mathbf{x}'=(\mathbf{x}'_1, \ldots, \mathbf{x}'_n), \mathbf{x}'_i\in \Delta_i,\not \exists \phi>0: \\
\forall i\in N, \mathbf{x}_i' \cdot U_i \cdot \prod_{j\neq i} \mathbf{x}_j' - \mathbf{x}_i \cdot U_i \cdot \prod_{j\neq i} \mathbf{x}_j \geq 0 \\
\exists i\in N, \mathbf{x}_i' \cdot U_i \cdot \prod_{j\neq i} \mathbf{x}_j' - \mathbf{x}_i \cdot U_i \cdot \prod_{j\neq i} \mathbf{x}_j \geq \phi
\label{def:weakPareto}
\end{multline}
\end{scriptsize}

We apply the same transformation $\tilde{U}_i = M_1 \cdot (\mathbf{x}_i \cdot  U_i \cdot  \prod_{j\neq i} \mathbf{x}_j)- U_i$ used in the proof of Theorem~\ref{thm:efficiency}. The problem of verifying whether $\mathbf{x}$ is strong Pareto efficient can be formulated in terms of optimization problem as follows. If there is no $\underline{i}\in N$ such that the optimal value $\gamma^*$ of the optimization problem:

\begin{scriptsize}
\begin{align}
\min_{\gamma, \mathbf{x}_i'~i\in N}\qquad	& \gamma		\label{eq:minmaxSN1} \\
						& 0							&& \geq \mathbf{x}_i' \cdot \tilde{U}_i \cdot \prod_{j\neq i} \mathbf{x}_j'	& \forall i\in N	\label{eq:minmaxSN2} \\
						& \gamma						&& \geq \mathbf{x}_i' \cdot \tilde{U}_{\underline{i}} \cdot \prod_{j\neq {\underline{i}}} \mathbf{x}_j'	& 		\label{eq:minmaxSN2} \\
						& \mathbf{1}^T\cdot \mathbf{x}'_i	&& = 1		& \forall i\in N							 \label{eq:minmaxSN3} \\
						& \mathbf{x}'_i					&& \geq \mathbf{0} & \forall i\in N						 \label{eq:minmaxSN4}
\end{align}
\end{scriptsize}

\noindent is strictly negative, then $\mathbf{x}_i$ is strongly Pareto efficient. The above optimization problem, as the problem~(\ref{eq:minmaxN1})--(\ref{eq:minmaxN4}), always admits optimal solution in which the size of the support of the players' strategies is upper bounded by $N$. It follows from the fact that, once the strategies of all the players except one have been fixed, the optimization problem is linear in the strategy of the remaining player and therefore there is a basic solution with at most $n$ non--zero variables. Thus, an optimal solution can be found by as we do in the proof of Theorem~\ref{thm:efficiency}, except that here we need to repeat the procedure for every $\underline{i} \in N$, requiring an extra cost that is linear in $n$, a except for a simple variation of the definition of formula $\phi$ that we need to decide. More precisely, formula $\psi$ is redefined as follows:

\begin{scriptsize}
\begin{multline*}
\hspace{-0.3cm}\psi(N,\underline{i},\mathbf{x},\mathsf{supp})=\\
\hspace{0.1cm}\left\{
\begin{array}{lr}
\left(\sum_{a_1\in \mathsf{supp}_1}\ldots\sum_{a_n\in \mathsf{supp}_n} (\tilde{U}_{\underline{i}}(a_1,\ldots,a_n)\cdot \prod_{k\in N}x'_{k,a_{k}}\leq0)\right) &\hspace{-0.2cm}\wedge\\
\bigwedge_{i\in N} \left(\sum_{a_1\in \mathsf{supp}_1}\ldots\sum_{a_n\in \mathsf{supp}_n} (\tilde{U}_i(a_1,\ldots,a_n)\cdot \prod_{k\in N}x'_{k,a_{k}}<0)\right) &\hspace{-0.2cm}\wedge\\
\bigwedge_{i\in N} \left(\sum_{a_i\in \mathsf{supp}_i}x'_{i,a_i}=1\right) &\hspace{-0.2cm}\wedge\\
\bigwedge_{i\in N, a_i\in \mathsf{supp}_i} \left(x'_{i,a_i}\geq 0\right)
\end{array}\right.
\end{multline*}
\end{scriptsize}

\noindent This concludes the proof.
\hfill$\Box$

\noindent From Theorem~\ref{thm:verificationSNE}, we can state the following.
\begin{cor}
Given an $n$--player game, the problem of finding an SNE is in~$\mathcal{NP}$ when $n$ is constant.
\label{cor:NP}
\end{cor}
Combining Corollary~\ref{cor:NP} with the hardness result presented in Corollary~5 of~\cite{sandholmComplexiy2008}, we can state the following.

\begin{cor}
Given an $n$--player game, the problem of finding an SNE is~$\mathcal{NP}$--complete when $n$ is constant.
\label{cor:NPcomplete}
\end{cor}

\noindent Corollary~\ref{cor:NPcomplete} pushes (unless $\mathcal{P}=\mathcal{NP}$) for the case in which the size of the smallest support of the SNEs rises in $\min_{i}\{m_i\}$; otherwise the existence of an SNE could be verified by enumerating a number of supports that is polynomial in the size of the game, thus requiring polynomial time. However, the corollary can be proven regardless of whether $\mathcal{P}=\mathcal{NP}$:

\begin{cor}
It is possible to construct $n$--agent game instances with $(m_1,\ldots,m_n)$ actions s.t. the size of the smallest support of the SNEs is $\Omega(\min_{i}\{m_i\})$.
\label{cor:largesupport}
\end{cor}

\noindent \emph{Proof}. It is possible to find instances in which the unique SNE has a support per agent with size $\frac{\min_{i}\{m_i\}}{2}$, see~\cite{appendice}.\hfill$\Box$

The above results can be easily extended to the case of $\varepsilon$--SNE. In order to verify whether a strategy profile $\mathbf{x}$ is weakly $\varepsilon$--Pareto efficient, it is enough to substitute $<0$ with $<\varepsilon$ in the first row of $\psi$. The verification of an $\varepsilon$--SNE is the same of SNE except that we need to verify weak $\varepsilon$--Pareto efficiency instead of weak Pareto efficiency. Thus, the same results in Theorems~\ref{thm:efficiency} and~\ref{thm:verificationSNE} and Corollaries~\ref{cor:NP}, \ref{cor:NPcomplete}, \ref{cor:largesupport} stated for SNE hold also for $\varepsilon$--SNE.


\section{SNE finding with 2 agents}
\label{sub:SNEcomputation}

With non--degenerate games, a simple algorithm to find an SNE is to enumerate the NEs and to verify whether there is an SNE. However, NE enumeration is $\#\mathcal{P}$--hard~\cite{sandholmComplexiy2008}, while SNE finding  is $\mathcal{NP}$--complete. Thus it should likely be possible to design more efficient algorithms. Here we explore alternative better approaches that also work when there is a continuum of equilibria.

\subsection{Basic algorithm}

Given the difficulties in deriving a finite set of (necessary and sufficient) constraints for the membership of a strategy to be in the Pareto frontier, we propose an algorithm that iterates between the computation of an NE and the verification of an SNE. Basically, the algorithm will compute an NE and verify whether or not it is an SNE.  If not, it will compute a new NE excluding the space of strategies that are dominated by strategy profile found during the SNE verification.  This process repeats until an SNE is found or it is proven that none exists. Our idea is supported by the experimental evaluation presented in~\cite{sandholmgilpinconitzer2005}, where the authors show that the compute time to find an NE (even optimal) is negligible w.r.t. the compute time to enumerate all the NEs, so calling an NE--finding oracle a number of times can be faster than enumerating all the NEs.

The algorithm we propose is essentially a \emph{spatial branch--and--bound} algorithm. For simplicity, we focus on the case with two agents. A state is denoted by $s$ and the set of states by $S$. Each state $s$ is associated with a convex subspace $V_s$ of the solution space of the problem~(\ref{gre_zero})--(\ref{sum_one}). Specifically, $V_s$ is defined by box constraints over $v_1$ and $v_2$:  $\underline{v}_{1,s} \leq v_1 \leq \overline{v}_{1,s}$ and $\underline{v}_{2,s} \leq v_2 \leq \overline{v}_{2,s}$ (where $\overline{v}_i$ and $\underline{v}_i$ are the upper and lower bounds respectively of the box).  We denote by $s_0$ the state in which $\underline{v}_i = \underline{U}_i$ and $\overline{v}_i = \overline{U}_{i}$ for every $i\in N$, where $\underline{U}_i$ and $\overline{U}_i$ are respectively the minimum and maximum entries of $U_i$. Any SNE, if exists, is in $V_{s_0}$.

\begin{algorithm}
\begin{algorithmic}[1]
\STATE $S\longleftarrow \mathsf{initialize}$
\REPEAT
	\STATE $s \longleftarrow \mathsf{remove}(S)$
        	\STATE $(\vartheta,\mathbf{x})\longleftarrow \mathsf{findNE}(s)$
        	\IF {$\vartheta = \mathsf{true}$}
		\STATE $(\vartheta, \cdot, \mathbf{x}' )\longleftarrow \mathsf{verifySNE}(\mathbf{x})$
		\IF {$\vartheta = \mathsf{true}$}
			\RETURN $\mathbf{x}$	
	        	\ENDIF
		\STATE $S=S\cup  \mathsf{branch}(s,\mathbf{x}')$
		\STATE $S\longleftarrow \mathsf{filter}(S)$
	\ENDIF
\UNTIL{$S$ is empty}
\RETURN $\mathsf{false}$
\end{algorithmic}
\caption{$\mathsf{findingSNE}$}
\label{alg:SNEfinding}
\end{algorithm}

The algorithm is reported in Algorithm~\ref{alg:SNEfinding} and works as follows. At Step~1, set $S$ is populated with the initial states. Then the algorithm repeats Steps~3--10 while $S$ is not empty. At Step~3 a state $s$ is removed from $S$ and at Step~4 an oracle is called to find an NE in subspace $V_s$. If there is such an NE $\mathbf{x}$, at Step~6 the algorithm verifies whether $\mathbf{x}$ is an SNE and, in the affirmative case, $\mathbf{x}$ is returned. In the negative case in which $\mathbf{x}$ is Pareto dominated by $\mathbf{x}'$, at Step~9 new states are generated from the current state $s$ in which the subspace dominated by $\mathbf{x}'$ is excluded. At Step~10, redundant states in $S$ are pruned.  We now discuss the subroutines that the algorithm calls in more detail.

$S\longleftarrow \mathsf{initialize}$: Generates the initial set $S$ of states. We consider two possible initializations. In the first (\emph{init1}), we assign $S=\{s_0\}$. In the second (\emph{init2}), we compute all the pure--strategy profiles that are resilient to pure--strategy multilateral deviations. This can be done in polynomial time by comparing agents' utilities provided by each outcome w.r.t. the agents' utilities provided by all the other outcomes. Call $X$ the set of pairs $(\hat{v}_{1,h},\hat{v}_{2,h})$ of the agents' expected utilities given by these strategy profiles. We generate a number of states to exclude the subspace dominated by the elements in $X$. Order the elements of $X$ in increasing order of $\hat{v}_{1,h}$ and call $\overline{h}$ the number of elements in $X$. The following states $s$ are generated: for every $h\in \{1,\ldots,\overline{h}-1\}$ a state $s$ is generated with $V_s=[\hat{v}_{1,h},\hat{v}_{1,h+1}]\times[\hat{v}_{2,h+1},\overline{U}_2]$; two additional states with zero--measure $V_s$ are generated, the first with $V_s=[\underline{U}_1,\hat{v}_{1,1}]\times[\hat{v}_{2,1},\hat{v}_{2,1}]$, the second with $V_s=[\hat{v}_{1,\overline{h}},\hat{v}_{1,\overline{h}}]\times[\underline{U}_2,\hat{v}_{2,\overline{h}}]$. With respect to \emph{init1}, \emph{init2} excludes some dominated subspaces avoiding the algorithm to search for an NE in such subspaces, but, on the other hand, it can introduce a large number of states requiring the algorithm to call the NE--finding oracle many more times.

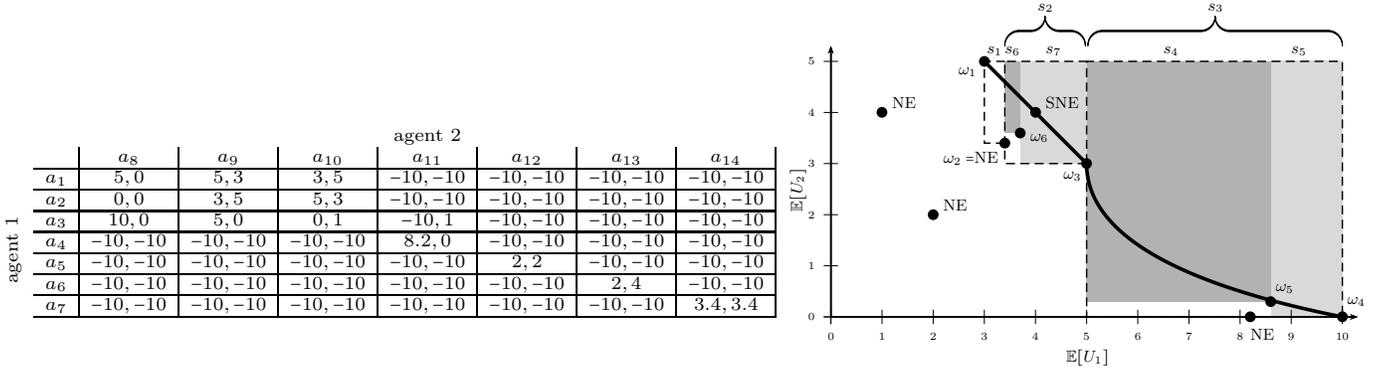
\begin{figure*}
\begin{minipage}{9.5cm}
\vspace{1.5cm}
\[
\begin{scriptsize}
\hspace{-0.3cm}
\begin{array}{rr|c|c|c|c|c|c|c|}
\multicolumn{2}{c}{}	&	\multicolumn{7}{c}{\textnormal{agent 2}} \\
	&		&	a_8	&	a_9	&	a_{10}&	a_{11}	&	a_{12}	&	a_{13}	&	a_{14}	\\ \cline{2-9}
\multirow{8}{*}{\begin{sideways}agent 1\end{sideways}}	&	
		a_1	&	5,0 		&	5,3		&	3,5		&	-10,-10	&	-10,-10	&	-10,-10	&	-10,-10	\\ \cline{2-9}
	&	a_2	&	0,0 		& 	3,5		&	5,3		&	-10,-10	&	-10,-10	&	-10,-10	&	-10,-10	\\ \cline{2-9}
	&	a_3	&	10,0 		& 	5,0		&	0,1		&	-10,1		&	-10,-10	&	-10,-10	&	-10,-10	\\ \cline{2-9}
	&	a_4	&	-10,-10	&	-10,-10 	& 	-10,-10	&	8.2,0		&	-10,-10	&	-10,-10	&	-10,-10	\\ \cline{2-9}
	&	a_5	&	-10,-10	&	-10,-10 	& 	-10,-10	&	-10,-10	&	2,2		&	-10,-10	&	-10,-10	\\ \cline{2-9}
	&	a_6	&	-10,-10 	&	-10,-10 	& 	-10,-10	&	-10,-10	&	-10,-10	&	2,4		&	-10,-10	\\ \cline{2-9}
	&	a_7	&	-10,-10	&	-10,-10 	& 	-10,-10	&	-10,-10	&	-10,-10	&	-10,-10	&	3.4,3.4		\\ \cline{2-9}
\end{array}
\end{scriptsize}
\]
\end{minipage}
\begin{minipage}{3.5cm}
\begin{pspicture}*(-1,-1)(10.3,5.3)
\newgray{lightgray}{0.85}
\newgray{gray}{0.7}
\scalebox{0.68}{

\pscustom[linestyle=none,fillstyle=solid,fillcolor=lightgray]{  \psline(8.6,0)(8.6,5)(10,5)(10,0) }
\pscustom[linestyle=none,fillstyle=solid,fillcolor=gray]{  \psline(8.6,0.30)(8.6,5)(5,5)(5,0.30) }
\pscustom[linestyle=none,fillstyle=solid,fillcolor=lightgray]{  \psline(3.7,3)(5,3)(5,5)(3.7,5) }
\pscustom[linestyle=none,fillstyle=solid,fillcolor=gray]{  \psline(3.4,3.6)(3.4,5)(3.7,5)(3.7,3.6) }

\psaxes[labelFontSize=\scriptstyle]{->}(0,0)(0,0)(10.3,5.3)


\psline[linecolor=black,linewidth=2pt]{-}(3,5)(5.000000,3.000000)(5.000500,2.960100)(5.002000,2.920400)(5.004500,2.880900)(5.008000,2.841600)(5.012500,2.802500)(5.018000,2.763600)(5.024500,2.724900)(5.032000,2.686400)(5.040500,2.648100)(5.050000,2.610000)(5.060500,2.572100)(5.072000,2.534400)(5.084500,2.496900)(5.098000,2.459600)(5.112500,2.422500)(5.128000,2.385600)(5.144500,2.348900)(5.162000,2.312400)(5.180500,2.276100)(5.200000,2.240000)(5.220500,2.204100)(5.242000,2.168400)(5.264500,2.132900)(5.288000,2.097600)(5.312500,2.062500)(5.338000,2.027600)(5.364500,1.992900)(5.392000,1.958400)(5.420500,1.924100)(5.450000,1.890000)(5.480500,1.856100)(5.512000,1.822400)(5.544500,1.788900)(5.578000,1.755600)(5.612500,1.722500)(5.648000,1.689600)(5.684500,1.656900)(5.722000,1.624400)(5.760500,1.592100)(5.800000,1.560000)(5.840500,1.528100)(5.882000,1.496400)(5.924500,1.464900)(5.968000,1.433600)(6.012500,1.402500)(6.058000,1.371600)(6.104500,1.340900)(6.152000,1.310400)(6.200500,1.280100)(6.250000,1.250000)(6.300500,1.220100)(6.352000,1.190400)(6.404500,1.160900)(6.458000,1.131600)(6.512500,1.102500)(6.568000,1.073600)(6.624500,1.044900)(6.682000,1.016400)(6.740500,0.988100)(6.800000,0.960000)(6.860500,0.932100)(6.922000,0.904400)(6.984500,0.876900)(7.048000,0.849600)(7.112500,0.822500)(7.178000,0.795600)(7.244500,0.768900)(7.312000,0.742400)(7.380500,0.716100)(7.450000,0.690000)(7.520500,0.664100)(7.592000,0.638400)(7.664500,0.612900)(7.738000,0.587600)(7.812500,0.562500)(7.888000,0.537600)(7.964500,0.512900)(8.042000,0.488400)(8.120500,0.464100)(8.200000,0.440000)(8.280500,0.416100)(8.362000,0.392400)(8.444500,0.368900)(8.528000,0.345600)(8.612500,0.322500)(8.698000,0.299600)(8.784500,0.276900)(8.872000,0.254400)(8.960500,0.232100)(9.050000,0.210000)(9.140500,0.188100)(9.232000,0.166400)(9.324500,0.144900)(9.418000,0.123600)(9.512500,0.102500)(9.608000,0.081600)(9.704500,0.060900)(9.802000,0.040400)(9.900500,0.020100)(10.000000,0.000000)

\uput{0}[0](4.2,4.2){SNE}
\uput{0}[0](2.2,2.2){NE}
\uput{0}[0](1.2,4.2){NE}
\uput{0}[0](2.2,3.1){$\omega_2=$NE}
\uput{0}[0](8.2,-0.33){NE}
\uput{0}[0](3.05,5.2){$s_1$}
\uput{0}[0](3.4,5.2){$s_6$}
\uput{0}[0](4.2,5.2){$s_7$}
\uput{0}[0](6.5,5.2){$s_4$}
\uput{0}[0](9,5.2){$s_5$}
\uput{0}[0](2.5,4.8){$\omega_1$}
\uput{0}[0](4.55,2.75){$\omega_3$}
\uput{0}[0](10.1,0.3){$\omega_4$}
\uput{0}[0](8.7,0.55){$\omega_5$}
\uput{0}[0](3.9,3.5){$\omega_6$}

\psbrace[ref=lt,rot=-90,nodesepB=-7pt,nodesepA=-4pt,braceWidth=0.75pt](4.975,5.3)(3.4,5.3){$s_2$}
\psbrace[ref=lt,rot=-90,nodesepB=-7pt,nodesepA=-4pt,braceWidth=0.75pt](10,5.3)(5.025,5.3){$s_3$}

\psdots[linecolor=black,dotsize=6pt](4,4)(8.2,0)(1,4)(2,2)(3.4,3.4)(10,0)(3,5)(5,3)(8.6,0.30)(3.7,3.6)

\uput{0}[0](4.6,-0.8){$\mathbb{E}[U_1]$}
\rput[tr]{90}(-0.8,2.9){$\mathbb{E}[U_2]$}

\psline[linestyle=dashed](5,0)(10,0)(10,5)(5,5)(5,0)
\psline[linestyle=dashed](5,3)(3.4,3)(3.4,5)(5,5)
\psline[linestyle=dashed](3.4,3.4)(3,3.4)(3,5)(3.4,5)

}
\end{pspicture}
\end{minipage}
\caption{Example of two--agent game (left) and its Pareto frontier (right).}
\label{fig:algorithm}
\end{figure*}

\begin{exam}
Consider the game reported in Fig.~\ref{fig:algorithm}. We show the composition of $X$ when \emph{init2} is used. $X$ is composed by four elements $\omega_1,\omega_2,\omega_3,\omega_4$. The generated states $s_1,s_2,s_3$ are reported in the figure as boxes delimited by dashed lines: $V_{s_1}=[3,3.4]\times[3.4,5]$,  $V_{s_2}=[3.4,5]\times[3,5]$, $V_{s_3}=[5,10]\times[0,5]$ (for simplicity we omit the states with zero--measure $V_s$ in what follows).
\end{exam}

$s \longleftarrow \mathsf{remove}(S)$: Receives the set of states $S$ as input, removes a state $s$ from $S$ according to some given strategy (e.g., depth first, breadth first, random), and returns $s$.

$(\vartheta,\mathbf{x})\longleftarrow \mathsf{findNE}(s)$: Receives a state $s$ as input, searches for an NE in the subspace bounded by $V_s$, and returns $\vartheta=\mathsf{true}$ and an NE $\mathbf{x}$, if there is an NE where agents' utilities are in $V_s$, and $\vartheta=\mathsf{false}$, otherwise. The introduction of constraints $V_s$ makes some algorithms available in the literature to find an NE not to be applicable, e.g., the Lemke--Howson. NE--finding oracles can be: PNS, LS--PNS~\cite{CeppiGattiPatriniRoccoAAMAS2010},  MIP Nash.

\begin{exam}
In Fig.~\ref{fig:algorithm}: $\mathsf{findNE}(s_1)$ would return the NE with $v_1=v_2=3.4$; $\mathsf{findNE}(s_2)$ could return either the NE returned with $s_1$ or the mixed NE with $v_1=v_2=4$; $\mathsf{findNE}(s_3)$ would return the NE with $v_1=8.2$ and $v_2=0$; $\mathsf{findNE}(s)$ with $s\in \{s_4,s_5,s_6\}$ would return no NE;  $\mathsf{findNE}(s_7)$ would return the mixed NE with $v_1=v_2=4$.
\end{exam}

$S' \longleftarrow \mathsf{branch}(s,\mathbf{x})$: Receives a state $s$ and a strategy profile $\mathbf{x}$ and returns a number of new states in which the subspace dominated by $\mathbf{x}$ is excluded. The generation of the new states is as follows. Call $\hat{v}_{1,s}$ and $\hat{v}_{2,s}$ the expected utilities of agents~1 and~2 respectively provided by $\mathbf{x}$ given as input. A new state $s'$ is generated with $V_{s'}= [\underline{v}_{1,s}, \min\{\overline{v}_{1,s},\hat{v}_{1,s}\}]\times[ \hat{v}_{2,s}, \overline{v}_{2,s}]$ and, if $\min\{\overline{v}_{1,s},\hat{v}_{1,s}\}\neq \overline{v}_{1,s}$, a new state $s''$  is generated with $V_{s''}=[\min\{\overline{v}_{1,s},\hat{v}_{1,s}\},\overline{v}_{1,s}]\times[\underline{v}_{2,s},\overline{v}_{2,s}]$.

\begin{exam}
In Fig.~\ref{fig:algorithm}: $\mathsf{branch}(s_3,\mathbf{x}_{\omega_5})$ produces two states, $s_4$ and $s_5$, where $V_{s_4}=[5,8.2]\times[0.4,5]$ (dark gray) and $V_{s_5}=[8.2,10]\times[0,5]$ (light gray); $\mathsf{branch}(s_2,\mathbf{x}_{\omega_6})$ produces two states, $s_6$ and $s_7$, where $V_{s_6}=[3.4,3.7]\times[3.6,5]$ (dark gray) and $V_{s_7}=[3.7,5]\times[3,5]$ (light gray). ($\mathbf{x}_{\omega_j}$ is the strategy profile associated with $\omega_j$.)
\end{exam}

$S'\longleftarrow \mathsf{filter}(S)$: Receives the set of states $S$ as input and returns a set $S'$ subset of $S$ after having pruned states. If there is a pair of states $s,s'$ with $s'\neq s$ and such that $\underline{v}_{1,s}\leq \underline{v}_{1,s'}$ and  $\underline{v}_{2,s}\leq \underline{v}_{2,s'}$, then we can remove $s$ and $s'$ and add a new state $s''$ with $V_{s''}=[\underline{v}_{1,s},\overline{v}_{1,s'}]\times[\underline{v}_{2,s'},\overline{v}_{2,s'}]$.

\begin{exam}
In Fig.~\ref{fig:algorithm}: suppose $S=\{s_1,s_4,s_5,s_6,s_7\}$. States $s_1$ and $s_6$ can be removed and state $s_8$ with $V_{s_8}=[3,3.7]\times[3.6,5]$ can be added.
\end{exam}

\begin{thm}
Algorithm~\ref{alg:SNEfinding} is sound and complete.
\end{thm}

\noindent Soundness is by definition of $\mathsf{findNE}$ and of $\mathsf{verifySNE}$. Completeness is due to $\mathsf{filter}$ and $\mathsf{branch}$ removing only Pareto-dominated solution subspaces. When NEs are finite, Algorithm~\ref{alg:SNEfinding} terminates in finite time. In the next section, we propose a variation able to deal also with games admitting a continuum of NEs (however, while  Algorithm~\ref{alg:SNEfinding} can be, in principle, extended to the case with more than two agents, it is not clear whether the algorithm described in the next section can be extended to such case).

\subsection{Iterated MIP 2StrongNash}
\label{subsec:IMIPSN}

By exploiting MILP tools we can avoid having to spatially branch the solution space into subspaces.  We call this new algorithm \emph{iterated MIP 2StrongNash}, given that it exploits iteratively a slightly modified version of MIP Nash for SNE. It is reported in Algorithm~\ref{alg:MIP2SN}. With this algorithm, we have a unique state at each iteration and, instead of generating additional states, we add non--convex constraints. The branch and bound process per state is relegated to the MILP solver.

\begin{algorithm}
\begin{algorithmic}[1]
\STATE $\mathsf{initialize}$
\WHILE{$\mathsf{true}$}
        	\STATE $(\vartheta,\mathbf{x})\longleftarrow \mathsf{findNE}$
        	\IF {$\vartheta = \mathsf{false}$}
		\RETURN $\mathsf{false}$
	\ENDIF
	\STATE $(\vartheta, \cdot, \mathbf{x}' )\longleftarrow \mathsf{verifySNE}(\mathbf{x})$
	\IF {$\vartheta = \mathsf{true}$}
			\RETURN $\mathbf{x}$	
	\ENDIF
	\STATE $\mathsf{branch}$
	\STATE $\mathsf{filter}$
\ENDWHILE
\end{algorithmic}
\caption{$\mathsf{Iterated~MIP~2StrongNash}$}
\label{alg:MIP2SN}
\end{algorithm}

The core of this algorithm is the MIP Nash formulation~\cite{sandholmgilpinconitzer2005} for the computation of an NE ($\mathsf{findNE}$):

\begin{scriptsize}
\begin{align}
\textnormal{constraints }&(\ref{gre_zero}),~(\ref{exp_utility}),~(\ref{sum_one}) \notag\\
\mathbf{1}v_i - U_{i}\cdot \mathbf{x}_{-i} & \leq \overline{U}_i\cdot (\mathbf{1}-\mathbf{b}_i) & \forall i\in N \label{MIPNash1}\\
\mathbf{x}_i & \leq \mathbf{b}_i	&	\forall i \in N \label{MIPNash2} \\
\mathbf{b}_i & \in \{0,1\}^{m_i}	& \forall i\in N \label{MIPNash3}
\end{align}
\end{scriptsize}

\noindent Given that the above program is a MILP, integer and/or linear constraints can be easily added. We now describe the subroutines employed in the algorithm.

$\mathsf{initialize}$. In the case of \emph{init1}, no additional constraint is added. In the case of \emph{init2}, after having found $X$, we add three constraints for each element $(\hat{v}_{1,k},\hat{v}_{2,k})$:

\begin{scriptsize}
\begin{align}
v_1	&	\geq 	\hat{v}_{1,k} +	(\underline{U}_1-\overline{U}_1)\cdot z_k			\label{branch1}\\
v_2	&	\geq	\hat{v}_{2,k} +	(\underline{U}_2-\overline{U}_2)\cdot (1-z_k)		\label{branch2}\\
z_k	&	\in \{0,1\}													\label{branch3}
\end{align}
\end{scriptsize}

\noindent where constraints (\ref{branch1}) and (\ref{branch2}) exclude that both $v_1$ and $v_2$ are simultaneously smaller than $\hat{v}_{1,k}$ and $\hat{v}_{2,k}$, respectively. A different binary variable $z_k$ is introduced for each~$k$.

$\mathsf{branch}$. After having found an NE $\mathbf{x}$ that is Pareto dominated by $\mathbf{x}'$ with agents' utilities $(\hat{v}_1,\hat{v}_2)$, constraints (\ref{branch1})--(\ref{branch2}) are added with a new binary variable $z_k$ and $\hat{v}_{1,k}=\hat{v}_1$ and $\hat{v}_{2,k}=\hat{v}_2$.

$\mathsf{filter}$. If there are two variables $z_k,z_{k'}$ with $z_k\neq z_{k'}$ and such that $\hat{v}_{1,k}\leq \hat{v}_{1,k'}$ and $\hat{v}_{2,k}\leq \hat{v}_{2,k'}$ then variable $z_k$ and the three constraints associated with $z_k$ can be removed.

It is easy to add features to the algorithm.  Here we discuss some relevant examples of that.

\emph{Social welfare maximization}. A way to find Pareto efficient solutions is to maximize the cumulative utility of the agents. We can do it by adding an objective function:

\begin{scriptsize}
\begin{align}
\max & \quad v_1+v_2
\end{align}
\end{scriptsize}

\noindent By employing this feature, Iterated MIP 2Strong--Nash returns an optimal SNE. This feature allows the algorithm to terminate within finite time even on degenerate games that admit a continuum of NEs. This is because, although a continuum of NEs contains infinitely many NEs, there is only one NE of the continuum that maximizes social welfare (in a NE continuum, the utility of only one agent can vary). Thus, either such an NE is an SNE and the algorithm terminates or all the NEs of the continuum are Pareto dominated and therefore they are all discarded in one iteration.


\emph{Upper bound over social welfare}. We can exploit the social welfare maximization also to add constraints over the solution space. Specifically, call $v^*$ the value of the objective function at iteration~$k$.  We can add the constraint

\begin{scriptsize}
\begin{align}
v_1+v_2	&	\leq \overline{sw} \label{con:upper}
\end{align}
\end{scriptsize}

\noindent at iteration $k+1$ where $\overline{sw}=v^*$. Therefore, $\overline{sw}$ decreases monotonically at each iteration.

\emph{Lower bound over social welfare}. We can find a linear lower bound over the social welfare as follows. We order all the constraints (\ref{branch1})--(\ref{branch3}) in increasing order of $\hat{v}_{1,k}$. Let $\overline{k}$ to be the number of constraints (\ref{branch1})--(\ref{branch3}). We define $\underline{sw}= \min_{k\in \{1,\ldots,\overline{k}-1\}}\{\hat{v}_{1,k},\hat{v}_{2,k+1}\}$. It is the lower bound on social welfare. If $\underline{sw}>\overline{sw}$, then there is no SNE. Otherwise, we can call $\mathsf{findNE}$ with an additional constraint

\begin{scriptsize}
\begin{align}
v_1+v_2	&	\geq \underline{sw} \label{con:lower}
\end{align}
\end{scriptsize}

\noindent Constraints (\ref{con:upper}) and (\ref{con:lower}) are redundant. However, they can be used by the MILP solver to speed up the compute time.


\section{Experimental evaluation}

We experimentally evaluate the various configurations of Iterated MIP 2StrongNash, and subsequently, due to limited space, we report our experimental observations obtained from a preliminary comparison of such an algorithm w.r.t. Algorithm~\ref{alg:SNEfinding} and w.r.t. a simple algorithm that combines NE enumeration with SNE verification. We considered four configurations of Iterated MIP 2StrongNash: $\mathsf{mode}~0$ in which \emph{init1} is used and no additional features are active, $\mathsf{mode}~1$ in which \emph{init1} is used with the social welfare maximization and the upper bound over the social welfare, $\mathsf{mode}~2$ is as $\mathsf{mode}~1$ plus \emph{init2}, and $\mathsf{mode}~3$ is as $\mathsf{mode}~2$ without social welfare maximization. Although we showed that it is possible to produce instances such that each configuration is the best and instances in which each configuration is the worst (omitted here due to reasons of space), it is interesting to evaluate them in the average case. We implemented the SNE verification in the C programming language and the NE--finding oracle in AMPL with CPLEX. We used an Intel 2.20GHz processor and Linux kernel 2.6.32.

\textbf{GAMUT and SNEs}. Turns out that the ubiquitous benchmark testbed for NE, GAMUT~\cite{gamut}, is not a suitable testbed for SNE finding. We generated 20 instances of 2--agent games per GAMUT game class with a number of actions per agent in $\{25,50\}$. We report in Tab.~\ref{tab:GAMUT50} the following data for 50 actions per agent: percentage of instances admitting an SNE (`Y'), percentage of instances admitting mixed--strategy SNEs (`mY'), percentage of instances admitting only mixed--strategy SNEs (`omY'), average compute time of Iterated MIP 2StrongNash with $\mathsf{mode}~0$ (`time'), average compute time when there is an SNE (`time|Y'), and average compute time where there is no SNE  (`time|N').

Only TravellerDilemma and BertrandOlygopoly instances never admit SNEs, whereas, with the other classes, the percentage of instances admitting an SNE is high on average (with many classes, all the generated instances admit SNEs). Interestingly, only WarOfAttrition instances admit mixed--strategy SNEs, but no instance admits only mixed--strategy SNEs. The results with 25 actions per agent are similar~\cite{appendice}. Therefore, with all the generated GAMUT instances an SNE can be found quickly by enumerating all the pure--strategy NEs and employing SNE verification.  The time spent by Iterated MIP 2Strong Nash to find an SNE is much longer than the time spent by MIP Nash to find an NE, see, e.g., \cite{sandholmgilpinconitzer2005}. (Computing an SNE requires about 100 times the time needed to find an NE.) As in the case of NE finding, the hardest classes are the Covariant, Graphical, Polymatrix, and Random. Finally, finding an SNE when it exists is significantly faster than certifying that it does not exist.

\textbf{Novel \emph{ad hoc} game instances}. We design a game  generator, named \emph{mixed--strategy SNE instance generator} (MISSING), able to produce instances admitting only mixed--strategy SNEs (see~\cite{appendice}). The parameters of the generator are: number of actions per agent, SNE existence or non--existence (in the case of existence, we have only one SNE), size of the support of the SNE (per agent), number of non--strong NEs  (generated randomly with a social welfare that can be larger or smaller than the SNE's one). We generated 20 instances per combination of the following parameters: $m_1=m_2=m$ from 10 to 100, $|\mathsf{supp}|\in\{0,2,4,6,8,10\}$ where $|\mathsf{supp}|=0$ means that there is no SNE, other NEs $\in\{0,2,4,6,8,10,12\}$. We solved these instances with Iterated MIP 2StrongNash. The main results are in Figs.~\ref{fig:time} and~\ref{fig:iteration}, complete results are in~\cite{appendice}: the black solid line is with no other NEs, the black dashed line is with 6 non--strong NEs, the gray line is with 12 non--strong NEs.

Initially, we evaluate the different features of the algorithm. We observed that the social welfare maximization leads to a significant reduction of the iterations number (about 3--4 times), but it strongly negatively affect the compute time per iteration spent by the NE--finding oracle (up to 24 times). Interestingly, we observed that the largest increase in NE--finding compute time is in the iteration in which the SNE is returned, while with the other iterations the increase is smaller. This is because the SNE presents a large support and it results harder to be found, instead the other non--strong NEs present a small support. We observed that the lower bound over the social welfare does not provide significant improvements with the used instances. We observed that \emph{init2} reduces the iterations number without increasing significantly the compute time per iteration of the NE--finding oracle. Thus, we focus on $\mathsf{mode}~2$ and $\mathsf{mode}~3$, being the most significative configurations.

It can be observed in the figures that $\mathsf{mode}$~3 has the best compute time (4 times better at $m=100$). Differently from $\mathsf{mode}~2$, $\mathsf{mode}~3$ does not present an exponential growth around $m=100$. Surprisingly, as the number of non--strong NEs increases, the compute time does not increase significantly. There are two main reasons. First, a large portion of time is devoted to the verification (50\%--80\% with $\mathsf{mode}$~3 and 10\%--30\% with $\mathsf{mode}$~2) and almost all the verification time is spent on the verification of the SNE (this time is the same for all the instances and modes), while the verification of the NEs that are not SNEs requires much less time. Second, the NE--finding oracle rapidly finds the NEs that are not SNEs (because they have a small support of 1 or 2), while finding the SNE (that has a much larger support) is much harder. Thus the presence of additional small--supported non--strong NEs produces non--significative effects. Instead, compute time rises exponentially in $|\mathsf{supp}|$ of SNE.  This shows the need for developing techniques to speed up the SNE verification algorithm. It shows also that in the average case $\mathsf{mode}$~3 is the best and this is because with $\mathsf{mode}$~2 the NE--finding oracle is much more expensive. We believe that it would be interesting to develop new game classes combining hard GAMUT classes with MISSING instances to evaluate for what classes $\mathsf{mode}$~3 would still be best and for what classes welfare maximizing is useful.

\begin{table}[h]
\begin{center}
\begin{tiny}
\begin{tabular}{|r|r|r|r|r|r|r|r|}
\hline
class               			&   Y 		&   \hspace{-0.1cm}mY\hspace{-0.1cm}		&  \hspace{-0.1cm}omY\hspace{-0.1cm}		& 	\hspace{-0.1cm}time\hspace{-0.1cm} 		&	 \hspace{-0.1cm}time|Y\hspace{-0.1cm}	&    \hspace{-0.1cm}time|N\hspace{-0.1cm} 	\\ \hline\hline
\hspace{-0.1cm}BertrandOlygopoly\hspace{-0.1cm}		& \hspace{-0.1cm}0\%\hspace{-0.1cm}		& 0\%		& 0\%		& 1963.6~s		&	--		& 1963.6~s		\\ \hline
\hspace{-0.1cm}BidirectionalLEG--RG\hspace{-0.1cm} 	& \hspace{-0.1cm}95\%\hspace{-0.1cm}		& 0\%		& 0\%		&  29.6~s 		&       29.6~s 	&  8.6~s   		\\ \hline
\hspace{-0.1cm}CovariantGame--Rand\hspace{-0.1cm}  	& \hspace{-0.1cm}55\%\hspace{-0.1cm}		& 0\%		& 0\%		&  \hspace{-0.2cm}17,701.5~s	&      53.9~s	&  \hspace{-0.2cm}39,270.9~s	\\ \hline
\hspace{-0.1cm}CovariantGame--Zero\hspace{-0.1cm}	& \hspace{-0.1cm}40\%\hspace{-0.1cm}		& 0\%		& 0\%		& \hspace{-0.2cm} 3,734.1~s 	&      \hspace{-0.2cm}1,184.2~s  &\hspace{-0.2cm} 5,434.1~s	\\ \hline
\hspace{-0.1cm}GraphicalGame--Road\hspace{-0.1cm} 	& \hspace{-0.1cm}35\%\hspace{-0.1cm}		& 0\%		& 0\%		& \hspace{-0.2cm}15,247.7~s	&	495.0~s	& \hspace{-0.2cm}23,191.4~s	\\ \hline
\hspace{-0.1cm}GraphicalGame--SG\hspace{-0.1cm} 	& \hspace{-0.1cm}60\%\hspace{-0.1cm}		& 0\%		& 0\%		& \hspace{-0.2cm}10,217.3~s	&      \hspace{-0.2cm}4,029.5~s	 &\hspace{-0.2cm} 19,498.9~s	\\ \hline
\hspace{-0.1cm}MinimumEffortGame\hspace{-0.1cm} 	& \hspace{-0.1cm}100\%\hspace{-0.1cm}		& 0\%		& 0\%		& 22.3~ s		&       22.3~s	& --			\\ \hline
\hspace{-0.1cm}PolymatrixGame--CG\hspace{-0.1cm}	& \hspace{-0.1cm}45\%\hspace{-0.1cm}		& 0\%		& 0\%		& \hspace{-0.2cm}3,860.0~s	&      892.9~s	& \hspace{-0.2cm} 6,287.6~s	\\ \hline
\hspace{-0.1cm}PolymatrixGame--SW\hspace{-0.1cm} 	& \hspace{-0.1cm}25\%\hspace{-0.1cm} 		& 0\% 		& 0\% 		& \hspace{-0.2cm}6,950.3~s	&     \hspace{-0.2cm} 3,478.7~s	&     \hspace{-0.2cm}8,107.5~s	\\ \hline
\hspace{-0.1cm}RandomGame\hspace{-0.1cm} 			& \hspace{-0.1cm}45\%\hspace{-0.1cm}		& 0\%		& 0\% 		&\hspace{-0.2cm} 6,998.0~s	&       \hspace{-0.2cm}6,229.9~s	&     \hspace{-0.2cm} 7,628.2~s	\\ \hline
\hspace{-0.1cm}TravelersDilemma\hspace{-0.1cm}  		& \hspace{-0.1cm}0\%\hspace{-0.1cm}		& 0\%		& 0\%		& 101.7~s		&	--		& 101.7~s		\\ \hline
\hspace{-0.1cm}UniformLEG--CG\hspace{-0.1cm} 		& \hspace{-0.1cm}75\%\hspace{-0.1cm}		& 0\%		& 0\%		& 33.2~s		& 	35.9~s	& 25.2~s		\\ \hline
\hspace{-0.1cm}UniformLEG--SG\hspace{-0.1cm}		& \hspace{-0.1cm}100\%\hspace{-0.1cm}		& 0\% 		& 0\% 		& 8.5~s 		& 8.5~s		& --			\\ \hline
\hspace{-0.1cm}WarOfAttrition\hspace{-0.1cm}			& \hspace{-0.1cm}100\%\hspace{-0.1cm}		& 35\%		& 0\%		&  13.8~s		&	13.8~s	&       --	 	\\ \hline
\end{tabular}
\end{tiny}
\end{center}
\caption{Results with 50x50 GAMUT instances: instances that admit at least one SNE (Y), instances that admit at least one mixed--strategy SNE (mY), instances that admit only mixed--strategy SNE (omY), compute time (time), compute time when there is an SNE (time|Y), compute time when there is no SNE (time|N).}
\label{tab:GAMUT50}
\end{table}

\textbf{Comparison to other algorithms}. We used MISSING instances in the following comparisons. 

We compared Iterated MIP 2StrongNash and Algorithm~\ref{alg:SNEfinding} when MIP~Nash is the NE--finding oracle in terms of the compute time. Iterated 2StrongNash dramatically outperforms Algorithm~\ref{alg:SNEfinding}---by more than one order of magnitude. The main reason is that Algorithm~\ref{alg:SNEfinding} solves each NE--finding subproblem on $V_s$ independently from the others and there is no a clear heuristic to select the next subproblem to solve. It is interesting to observe that the relationship between Iterated MIP 2StrongNash and Algorithm~\ref{alg:SNEfinding} with MIP Nash as the NE--finding oracle is close to the relationship between MIP Nash and PNS. MIP Nash deals with all the supports together, while PNS works with each support separately. PNS adopts a specific heuristic, scanning the supports from smallest to largest (this holds in almost all the GAMUT instances). When there is no NE with small support, PNS is dramatically worse than MIP Nash. In our case, we did not find a  good heuristic for Algorithm~\ref{alg:SNEfinding} because for each heuristic it is possible to design a worst--case game instance in which the number of calls to the NE--finding oracle is equal to the number of NEs. Thus, we selected randomly the next state $s$ to which apply the NE--finding oracle.

The adoption of PNS as the NE--finding oracle in Algorithm~\ref{alg:SNEfinding} is not satisfactory. It beats Iterated MIP 2StrongNash only when Algorithm~\ref{alg:SNEfinding} finds an SNEs by a sequence of NE--finding subproblems, each admitting an NE with small support. Otherwise, if some subproblem does not admit any NE or admits one with large support, PNS must enumerate a very large number of supports, making it slow.

Finally, we report experimental results on the employment of SNE verification with an NE enumeration algorithm~\cite{avis} that is stopped every time an NE is found and then verified. Enumerating all the NEs, necessary when an SNE does not exist, requires more than 300~s with 20x20 games and does not terminate in one day with 50x50 games, while Iterated MIP 2StrongNash terminates within 150~s with 100x100 games. Even when an SNE exists, we did not observe any termination in one day with 100x100 games.

\section{Conclusions and future work}

In this paper, we studied the verification (with $n$ agents) and computation (with two agents) of a strong Nash equilibrium (SNE). A number of results for Nash equilibrium (NE) are known, but that concept is inappropriate when coalitions are an issue. Unlike in NE finding, here mixed--strategy deviations must be taken into account, which makes the problem significantly more difficult. We showed that the verification problem is nevertheless in $\mathcal{P}$ given an arbitrary $n$--agent game, and therefore the problem of finding an SNE is in $\mathcal{NP}$. Then, we exploited our algorithm for SNE verification to design an algorithm for  finding an SNE. It is based on spatial branch--and--bound and iterates between a NE--finding oracle and the verification algorithm. Finally, we experimentally evaluated our algorithm. We showed that the instances from the ubiquitous NE benchmark testbed, GAMUT, are not suitable for testing SNE--finding algorithms because all the instances either admit pure SNEs or do not admit any SNE. Then we compared different configurations of our algorithm using a new instance generator to identify the best one. With these instances, it turns out that (our algorithm for) SNE finding takes about 100 times as long as NE finding.  In addition, our algorithm dramatically outperforms the approach of enumerating NEs and verifying them for SNE; therefore it is the current state of the art for finding mixed--strategy SNEs even when NEs are finite.

\begin{figure}[t]
\hspace{-0.1cm}\begin{minipage}{4.1cm}

\begin{pspicture}*(-0.7,-1.0)(11,2)
\newgray{lightgray}{0.85}
\newgray{gray}{0.7}
\psset{xunit=0.035,yunit=.005}


\psaxes[Dx=20,Dy=60,labelFontSize=\scriptstyle]{->}(0,0)(0,0)(105,365)

\psline[linewidth=1pt]{-}(20,8.330000)(25,7.803750)(30,8.171250)(35,7.752500)(40,11.935000)(45,13.291250)(50,10.296250)(55,13.820000)(60,14.522500)(65,18.896250)(70,26.818750)(75,42.773750)(80,60.545000)(85,56.280000)(90,56.582500)(95,63.195000)(100,71.250000)

\psline[linestyle=dashed,linewidth=1pt]{-}(35,12.523750)(40,15.060000)(45,18.670000)(50,30.762500)(55,31.076250)(60,28.717500)(65,36.562500)(70,58.512500)(75,65.713750)(80,40.357500)(85,84.077500)(90,69.283750)(95,99.953750)(100,66.998750)

\psline[linecolor=gray,linewidth=1pt]{-}(50,33.272500)(55,43.331250)(60,39.225000)(65,65.963750)(70,50.597500)(75,83.636250)(80,58.173750)(85,65.422500)(90,53.912500)(95,89.291250)(100,121.136250)

\uput{0}[0](12,340){$\mathsf{mode}=3~/~\mathsf{supp}=10$}

\uput{0}[0](34,-140){$m_1=m_2$}

\end{pspicture}

\end{minipage}
\begin{minipage}{4cm}

\begin{pspicture}*(-0.7,-1.0)(11,2)
\newgray{lightgray}{0.85}
\newgray{gray}{0.7}
\psset{xunit=0.035,yunit=.005}

\psaxes[Dx=20,Dy=60,labelFontSize=\scriptstyle]{->}(0,0)(0,0)(105,365)

\psline[linewidth=1pt]{-}(20,10.057500)(25,7.567500)(30,8.733750)(35,11.485000)(40,12.546250)(45,19.418750)(50,32.641250)(55,56.466250)(60,51.622500)(65,65.307500)(70,85.553750)(75,105.523750)(80,121.165000)(85,139.177500)(90,235.905000)(95,228.961250)(100,165.182500)

\psline[linestyle=dashed,linewidth=1pt]{-}(35,38.627500)(35,38.800000)(40,40.538750)(45,32.723750)(50,47.958750)(55,52.227500)(60,69.733750)(65,84.241250)(70,66.016250)(75,85.340000)(80,108.903750)(85,140.516250)(90,129.382500)(95,181.895000)(100,322.271250)

\psline[linecolor=gray,linewidth=1pt]{-}(50,33.575000)(55,41.811250)(60,62.896250)(65,58.718750)(70,68.585000)(75,92.338750)(80,118.978750)(85,122.341250)(90,153.423750)(95,201.767500)(100,165.007500)

\uput{0}[0](12,340){$\mathsf{mode}=2~/~\mathsf{supp}=10$}

\uput{0}[0](34,-140){$m_1=m_2$}

\end{pspicture}

\end{minipage}
\caption{Average compute time (s).}
\label{fig:time}
\end{figure}
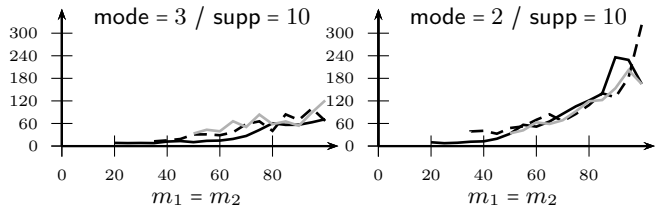

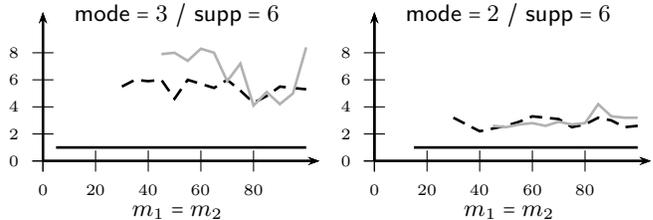
\begin{figure}[t]
\hspace{-0.2cm}\begin{minipage}{4.3cm}

\begin{pspicture}*(-0.6,-1.5)(11,2.2)
\newgray{lightgray}{0.85}
\newgray{gray}{0.7}
\psset{xunit=0.035,yunit=0.18}


\psaxes[Dx=20,Dy=2,labelFontSize=\scriptstyle]{->}(0,0)(0,0)(105,11)

\psline[linewidth=1pt]{-}(5,1.000000)(10,1.000000)(15,1.000000)(20,1.000000)(25,1.000000)(30,1.000000)(35,1.000000)(40,1.000000)(45,1.000000)(50,1.000000)(55,1.000000)(60,1.000000)(65,1.000000)(70,1.000000)(75,1.000000)(80,1.000000)(85,1.000000)(90,1.000000)(95,1.000000)(100,1.000000)

\psline[linestyle=dashed,linewidth=1pt]{-}(30,5.500000)(35,6.000000)(40,5.900000)(45,6.000000)(50,4.600000)(55,6.000000)(60,5.700000)(65,5.400000)(70,6.000000)(75,5.200000)(80,4.300000)(85,4.800000)(90,5.500000)(95,5.400000)(100,5.300000)

\psline[linecolor=gray,linewidth=1pt]{-}(45,7.900000)(50,8.000000)(55,7.400000)(60,8.300000)(65,8.000000)(70,5.900000)(75,7.200000)(80,4.100000)(85,5.100000)(90,4.200000)(95,5.000000)(100,8.400000)

\uput{0}[0](12,10.8){$\mathsf{mode}=3~/~\mathsf{supp}=6$}

\uput{0}[0](34,-3.8){$m_1=m_2$}

\end{pspicture}
\end{minipage}
\begin{minipage}{4cm}

\begin{pspicture}*(-0.6,-1.5)(11,2.2)
\newgray{lightgray}{0.85}
\newgray{gray}{0.7}
\psset{xunit=0.035,yunit=0.18}


\psaxes[Dx=20,Dy=2,labelFontSize=\scriptstyle]{->}(0,0)(0,0)(105,11)

\psline[linewidth=1pt]{-}(15,1.000000)(20,1.000000)(25,1.000000)(30,1.000000)(35,1.000000)(40,1.000000)(45,1.000000)(50,1.000000)(55,1.000000)(60,1.000000)(65,1.000000)(70,1.000000)(75,1.000000)(80,1.000000)(85,1.000000)(90,1.000000)(95,1.000000)(100,1.000000)

\psline[linestyle=dashed,linewidth=1pt]{-}(30,3.200000)(35,2.700000)(40,2.200000)(45,2.400000)(50,2.600000)(55,2.900000)(60,3.300000)(65,3.200000)(70,3.100000)(75,2.500000)(80,2.700000)(85,3.200000)(90,3.000000)(95,2.500000)(100,2.600000)

\psline[linecolor=gray,linewidth=1pt]{-}(45,2.600000)(50,2.500000)(55,2.700000)(60,2.800000)(65,2.600000)(70,2.900000)(75,2.700000)(80,2.800000)(85,4.200000)(90,3.300000)(95,3.200000)(100,3.200000)

\uput{0}[0](34,-3.8){$m_1=m_2$}

\uput{0}[0](12,10.8){$\mathsf{mode}=2~/~\mathsf{supp}=6$}

\end{pspicture}
\end{minipage}
\caption{Average iterations.}
\label{fig:iteration}
\end{figure}

Remaining open questions include the following:

\begin{ques}
How can Algorithms~\ref{alg:SNEfinding} and~\ref{alg:MIP2SN} be extended to the case with an arbitrary number of agents~\cite{AAAI2013}?
\end{ques}

\begin{ques}
Given a game with two agents and with $m$ actions per agent, what is the worst case number of calls to the NE--finding oracle?
\end{ques}

\begin{ques}
Given a game with an arbitrary number of agents, is it possible to formulate the SNE--finding problem with a finite set of (necessary and sufficient) constraints?
\end{ques}

\begin{ques}
What is the smoothed complexity of finding an SNE~\cite{IJCAI2013}?
\end{ques}

\section*{Acknowledgments}
The authors thank Zongque Xu for his collaboration. Nicola Gatti was supported by the Italian Ministry for Instruction, University and Research (MIUR) under Protocol N. 2009BZM837 (PRIN PeopleNet). Tuomas Sandholm was supported by the National Science Foundation under grants IIS--0964579, CCF--1101668, and IIP--1127832.

\bibliographystyle{abbrv}
\bibliography{ijcai11}

\end{document}